\documentclass[preprint,12pt]{elsarticle}
\usepackage{amssymb}
\usepackage{amsmath}
\begin{document}
\begin{frontmatter}
\title{Solvent Mixing Effect on Free-Energy Barrier and Stability for Molecular Recognition Driven by the Translational Motion of Solvent Molecules}

\author[label1,label2]{Mika Matsuo}
\author[label2]{Ryo Akiyama\corref{cor1}}
\cortext[cor1]{rakiyama@chem.kyushu-univ.jp}
\affiliation[label1]{organization={Faculty of Arts and Science},
             addressline={Kyushu University},
             city={Fukuoka},
             postcode={819-0395},
             country={Japan}}
\affiliation[label2]{organization={Department of Chemistry, Faculty of Science},
             addressline={Kyushu University},
             city={Fukuoka},
             postcode={819-0395},
             country={Japan}}

%% Abstract
\begin{abstract}
%% Text of abstract
We calculated the potentials of mean force (PMFs) between a ring-like host and a spherical guest in a solvent mixture. We adopted hard-body interactions between particles to discuss the effects of solvent-particle translational motion. The PMFs were obtained using the three-dimensional Ornstein--Zernike equation coupled with the modified hypernetted-chain closure (3D-MHNC--OZ theory). The entropic stabilization at the recognition site is confirmed, and a free-energy barrier wall is observed surrounding it. The free-energy barrier for the solvent mixture is much lower than that for the one-component solvent. Similar barrier reduction for the association of two spherical solute molecules has also been reported, and the mixing effect also reduced the dimerization stability. By contrast, the mixing effect does not significantly reduce recognition stability in the present study. In this respect, the behavior of the host--guest association is different from that of the association of two spherical solute molecules. 
\end{abstract}

%%Graphical abstract
%\begin{graphicalabstract}
%\includegraphics{grabs}
%\end{graphicalabstract}

%%Research highlights
%\begin{highlights}
%\item Research highlight 1
%\item Research highlight 2
%\end{highlights}

%% Keywords
\end{frontmatter}

\section{Introduction}
The host molecule combines with a specific guest molecule \cite{cram1985,cram1988,kyba1977,schneider2009,ohtaka2003,inoue1993,rekharsky1998,stites1997,fujisawa2004,kardos2004,gibb2012}. The process is often referred to as molecular recognition. Molecular recognition is usually found in a liquid phase that consists of solvent, host, and guest molecules. Many studies have focused only on the direct intermolecular interaction between host and guest molecules \cite{wipff1982-1,kollman1985-1,maye1991-1,lifson1983-1}. However, it is ultimately the thermodynamic quantities of the whole system, including solvent molecules, that determine whether molecular recognition will occur. The experimental results suggest that energy-driven molecular recognition is uncommon \cite{ohtaka2003,inoue1993,rekharsky1998,stites1997,fujisawa2004,kardos2004,gibb2012}. There are not many experiments that separate the association free energy of molecular recognition into entropy and energy (enthalpy) terms. Here, we introduce some experimental results for molecular recognition between HIV protease and indinavir \cite{ohtaka2003}. The free-energy differences between the initial separation and the final recognized states were obtained experimentally. The differences were divided into enthalpy and entropy contributions. In several cases, the enthalpy change was positive, and the molecular recognition was driven by entropy gain. This example is not an exception \cite{stites1997,fujisawa2004,kardos2004}. Entropy-driven molecular recognition has been reported experimentally; therefore, most of the explanations that are based only on direct attraction are contradicted by thermodynamic experimental results if they exist. In particular, entropic gains point to the importance of observing the translational motion of solvent molecules 
\cite{kinoshita2013,kinoshita2002,kinoshitaoguni2002,amano2010,amano2011}.

Entropic gain originating from the translational motion of the solvent molecules is often explained using Asakura--Oosawa theory \cite{asakura1954, asakura1958, vrij1976}, which is introduced using the simplest model system consisting of two large hard spheres immersed in a small hard-sphere fluid. The presence of a large sphere in small spheres forming the solvent generates a space from which the centers of the small spheres are excluded. The diameter of the excluded volume is the sum of the diameters of a large sphere and of a small sphere. Thus, the volume of the excluded volume is larger than the volume of a large sphere. When the large spheres are in contact, the excluded volumes for the small spheres overlap, and the volume available to the translational motion of the small spheres increases. This leads to an entropic gain for small spheres and induces an interaction between the large spheres. This effective interaction is known as the depletion interaction. The interpretation described above was given by Asakura and Oosawa in 1954 \cite{asakura1954, asakura1958, vrij1976}.

In the case of nonspherical solutes, the shape dependence of the depletion interaction is very interesting because it gives the guest selectivity by a host molecule. Kinoshita studied the depletion interaction between a hard body with a hemispherical cavity and a large hard sphere using the three-dimensional Ornstein--Zernike (OZ) equation coupled with the hypernetted-chain (HNC) closure (3D-HNC--OZ theory) \cite{kinoshita2002,kinoshitaoguni2002,amano2010,amano2011}. In that study, the selectivity of the diameter of the large sphere was shown. This lock-and-key relation can be explained using Asakura--Oosawa (AO) theory. In AO theory, as the overlap of the excluded volumes increases, the entropy gain increases. Here, the overlap is maximum when the large sphere fits exactly into the cavity. Therefore, the entropy gain becomes maximum. This selectivity could play an important role in the molecular recognition phenomenon in biological and synthetic systems \cite{akiyama2010,harano2005,kinoshita2009-1,imai2007,yoshidome2008,yasuda2011,du2012,oshima2011,yasuda2012,hayashi2014,hayashi2018}.

There is a problem with the shape of the effective interaction obtained by Asakura--Oosawa's idea, although the idea has an advantage in discussions of selectivity. For example, there are problems with estimating barrier shapes and heights during the recognition process. The remarkable simplification of the theory lies in modeling small spheres, namely depletants, as an ideal gas. The interaction between small particles is ignored in AO theory. Therefore, the small spheres do not construct any liquid structure near the solute molecules even when the packing fraction is large, such as 0.4. The effective interaction does not have oscillation induced by the liquid structure in this simplification. However, when the fluid of small spheres becomes dense, we can expect that the effective interaction is oscillatory due to the inhomogeneity of the number density of small spheres, namely the liquid structure. In particular, the high density of small spheres on the surface of large particles leads to a free-energy barrier to overcome before reaching the contact because the packing fraction of small spheres becomes high. Kinoshita et al. used integral equation theories to clarify the problem in the shape of the effective interaction \cite{kinoshita2002,kinoshitaoguni2002,amano2010,amano2011}.

Some studies show the depletion interaction between spherical solutes immersed in a hard-sphere mixture because the solvation structure strongly affects the shape of the effective interaction \cite{akiyama2006,karino2009}. The effective interaction depends on the mixing ratio, the size ratio, and the packing fraction. For example, as the number density of smaller spheres increases, the stability at contact increases, and the free-energy barrier becomes higher. Roth and Kinoshita reported a reduction in the free-energy barrier between two large spheres in multicomponent systems of smaller spheres \cite{roth2006}. The oscillatory structure of the effective interaction in a multicomponent system is less than that in a one-component system, and the free-energy barrier for the association process is significantly reduced in a multicomponent system. They adopted two large spheres as solute molecules. It seems that the reduction is due to interference from static density waves in the spatial distribution generated by various-sized particles. This idea of barrier reduction can be applied to reducing the recognition barrier in molecular recognition phenomena. 

In the present study, we conducted a theoretical study of molecular recognition using a spherical guest and a ring-like host molecule. The  host is a simple model of cyclodextrin known to exhibit size-selective recognition \cite{lostsson1996}. The potentials of mean force (PMFs) were calculated between the ring-like host molecule and the spherical guest molecule in a multicomponent mixture using the three-dimensional OZ equation coupled with the modified hypernetted-chain closure (3D-MHNC--OZ theory). Our previous study shows that MHNC theory is also a very accurate approximation for a system consisting of a fused hard-sphere solute with a fluid of hard spheres \cite{matsuo2024}. Therefore, MHNC theory is appropriate for the current study. The PMF results obtained with 3D-HNC--OZ and AO theory are also shown for comparison.

\section{Model and Method}
\subsection{Model}
\begin{figure}[h]
  \centering
  \includegraphics[width=0.45\columnwidth]{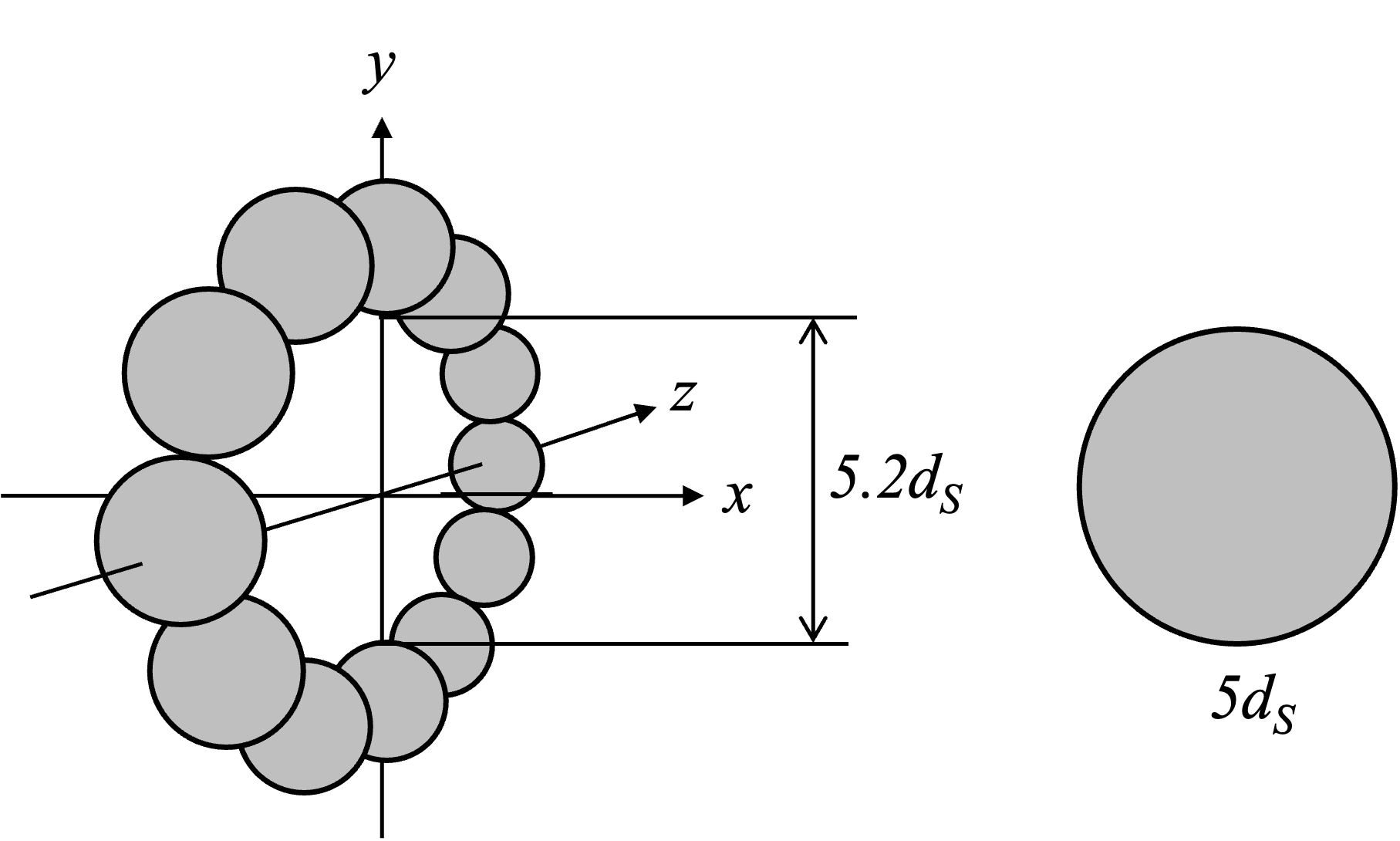}
 \caption{Host and guest molecules. Solute 1 is a spherical guest with a diameter of $5d_S$, which is slightly smaller than the host hole. Solute 2 is a ring-like host molecule, a fused hard sphere. It consists of $12$ hard spheres, and the diameter of each hard sphere is $2d_S$. The diameter of the central hole is $5.2d_S$. The center is located at the origin of the coordinate system.}
 \label{model}
\end{figure}
In the present study, we use hard-body interactions between particles to discuss the effective interaction between a guest and a host arising from packing effects. The solvent is a mixture of hard spheres. The guest (Solute 1) and host (Solute 2) are immersed in the solvent. They are shown in Fig. \ref{model}. The guest is a hard sphere with a diameter of $5d_S$. By contrast, the host (Solute 2) is a ring-like fused hard sphere consisting of 12 hard spheres, each with a diameter of $2d_S$. The centers of the spheres lie on a circle with a diameter of $7.2d_S$. Thus, the inner diameter of the host is approximately $5.2d_S$. Therefore, a hard sphere with a diameter of $5d_S$ can pass through the ring.

We adopted two solution models in the present study. One is a model in which guest particles ($i=1$) are included in the solution. We use this model in ``Section 3.1 Size Selectivity.'' In Section 3.1, we will discuss whether the guest particle with a diameter of $5d_S$ is valid as a guest particle recognized by the host molecule. Therefore, we will immerse the host molecule in a binary hard-sphere mixture containing low concentrations of guest particles and discuss the association stability. The two-component hard-sphere mixture is as follows. The number density of the guest particles is $\rho_1 =1.0\times 10^{-3}d_1^{-3}$, and the total packing fraction is kept at $0.380$. This packing fraction is virtually the same as that of ambient water. We examined five sizes of candidate guest particles (i.e., $1d_S$, $2d_S$, $3d_S$, $4d_S$, and $5d_S$). The diameter of the solvent particle is $d_S$.

Next, we introduce the second solution model. The fluid is a multicomponent mixture of hard spheres with the diameter $\lambda d_S$ in the range of $1<\lambda<4$. So, it does not include the guest with a diameter of $5d_S$. The total packing fraction of the mixture is kept at $0.380$ \cite{rothkinoshita2006,akiyama2006}. In another model, the host molecule and the guest particle with a diameter of $5d_S$ ($i=1, 2$) are at infinite dilution. We adopt this model when calculating the PMF between host and guest molecules in Sections 3.2 and 3.3. We examine nine mixtures (See Table \ref{tab1}). These mixtures have been examined in studies of effective interaction between two large hard spheres \cite{kinoshita1996, roth2006,rothkinoshita2006}. 
\begin{table}[]
  \caption{The nine systems examined. Systems 1--4 are one-component hard-sphere fluids; the others are mixtures. The total packing fraction remains constant at $0.380$ for all systems. For example, in System 9, the packing fractions of the small spheres with diameters $d_S$, $2d_S$, $3d_S$, and $4d_S$ are $0.095$, $0.095$, $0.095$, and $0.095$, respectively.}
  \label{tab1}
  \centering
  \begin{tabular}{ccccc}
    \hline
    System & $d_S$ & $2d_S$ & $3d_S$ & $4d_S$\\
    \hline \hline
    1 & 0.380 & $\cdots$ & $\cdots$ & $\cdots$ \\
    2 & $\cdots$ & 0.380 & $\cdots$ & $\cdots$ \\
    3 & $\cdots$ & $\cdots$ & 0.380 & $\cdots$ \\
    4 & $\cdots$ & $\cdots$ & $\cdots$ & 0.380 \\
    5 & 0.190 & 0.190 & $\cdots$ & $\cdots$ \\
    6 & 0.190 & $\cdots$ & 0.190 & $\cdots$ \\
    7 & 0.190 & $\cdots$ & $\cdots$ & 0.190 \\
    8 & 0.127 & 0.127 & 0.126 & $\cdots$ \\
    9 & 0.095 & 0.095 & 0.095 & 0.095 \\
    \hline
  \end{tabular}
\end{table}

\subsection[Integral Equation Theory]{Integral equation theory}
Taking computational cost into account, we adopt integral equation theory to obtain the correlation functions and evaluate the affinity between the host and guest in the present study. This decision is because evaluating various concentration cases in the fluids of a hard-sphere mixture with large size ratios using simulations requires considerable computational costs. By contrast, integral equation theory enables calculations for an infinite bulk system with very small computational cost.

The OZ equation is solved together with a closure equation \cite{hansen1985,mcquarrie1976}. For an $m$-component solvent in the presence of solute particles at infinite dilution, the OZ equation is written as
\begin{equation}
\label{OZ}
h_{ij}(r)=c_{ij}(r)+ \sum_{k=1}^{m+2} \rho_{k} \int c_{ik}(\mathbf{r}')h_{kj}(|\mathbf{r}-\mathbf{r}'|) d\mathbf{r}',
\end{equation}
where $\rho_k$ is the number density of species $k$, and $h_{ij}$ and $c_{ij}$ are the total and direct correlation functions, respectively. Indices $i$, $j$, and $k$ label particle species with $i=1, 2$ for solutes and $i=3,\dots,m+2$ for $m$ solvent species.  $r$ is the distance between the centers of the particles. The vectors are ${\mathbf{r}}=(x,y,z)$ and ${\mathbf{r'}}=(x',y',z')$. For a spherical system, that is, coupled with Eq. (\ref{OZ}), $\mathbf{r}$ is replaced with $r$. For a nonspherical system, the OZ equation is given on a 3D grid as follows:
\begin{equation}
\label{3D-OZ}
    h_{ij}(x,y,z)=c_{ij}(x,y,z)+ \sum_{k=1}^{m+2}\rho_{k} \int c_{ik}(x',y',z')h_{kj}(|\mathbf{r}-\mathbf{r}'|)dx'dy'dz'.
\end{equation}
Eq. (\ref{3D-OZ}) in the wavevector $\mathbf{k}$ space is written as:
\begin{equation}
\label{fourier}
\hat{\gamma}_{ij}(k_x,k_y,k_z)=\sum_{k=1}^{m+2}\rho_{k}  \hat{c}_{ik}(k_x,k_y,k_z)\hat{h}_{kj}(|\mathbf{k}|),
\end{equation}
where $\gamma=h-c$, the symbol ``\textasciicircum'' indicates the Fourier transform of the function of $\mathbf{r}$. The vector $\mathbf{k}=(k_x,k_y,k_z)$. $\hat{h}_{kj}(|\mathbf{k}|)$ is calculated using integral equation theory for the solvent.

In this study, effective interactions and three-dimensional spatial distribution functions are calculated based on these OZ equations with a closure relation. When we calculate the infinite dilution limit of component $i$, $\rho_i$ is set to 0. For example, in Figs. \ref{selective} and \ref{pxz}, when solving Eq. (\ref{OZ}), the number density of the host ($i=2$) is set to $\rho_2 = 0$. By contrast, for the guest ($i=1$), $\rho_1 =1.0\times 10^{-3}d_1^{-3}$. When we solve Eq. (\ref{3D-OZ}), which is a three-dimensional equation, after solving Eq. (\ref{OZ}), host ($i=2$) is introduced into the closure as an external field. Then, we obtain the three-dimensional spatial distribution functions. 

Here, I also show another example. In Figs. \ref{pz1},\ref{minimummap},\ref{minimummap4},\ref{pz2}, and \ref{pz3}, when solving Eq. (\ref{OZ}), $\rho_i = 0$ ($i=1, 2$). We obtain the correlation functions between solvent particles using Eq. (\ref{OZ}) with a closure relation. When solving Eq. (\ref{3D-OZ}) with a closure using the calculated correlation functions between solvent particles using Eq. (\ref{OZ}) with a closure relation, both host and guest ($i=1, 2$) are introduced into the closure as external fields. We obtain the effective interactions between the host and the guest in Figs. \ref{pz1},\ref{pz2} and \ref{pz3}, and the three-dimensional spatial distribution functions around the host and the guest in Figs. \ref{minimummap} and \ref{minimummap4}.

The closure relation is as follows:
\begin{equation}
\label{closure}
    c_{ij}(\mathbf{r})=\exp[-\beta u_{ij}(\mathbf{r})]\exp[\gamma_{ij}(\mathbf{r})+b_{ij}(\mathbf{r})]-\gamma_{ij}(\mathbf{r})-1,
\end{equation}
where $\beta=(k_BT)^{-1}$, $k_B$ is the Boltzmann constant and $T$ is the absolute temperature. The functions $u$ and $b$ are the interaction between particles and the bridge function, respectively. 

In the present study, the HNC closures ($b_{ij}(\mathbf{r})=0$) and modified closures with a bridge function proposed by Kinoshita\cite{MHNC2002, MHNC2003, nakamura2019,kinoshita2017} were used. The bridge function of the MHNC closure is as follows:
\begin{equation}
\label{MHNC}
    \begin{split}
        b_{ij}(\mathbf{r})&=-0.5\frac{\gamma_{ij}^2(\mathbf{r})}{1+0.8\gamma_{ij}(\mathbf{r})}\;\;(\gamma>0)\\
        &=-0.5\frac{\gamma_{ij}^2(\mathbf{r})}{1-0.8\gamma_{ij}(\mathbf{r})}\;\;(\gamma<0).
    \end{split}
\end{equation}
The reliability of the MHNC closure has been demonstrated for mixtures with large size asymmetry \cite{nakamura2019}. It has been verified that the MHNC closure provides more accurate spatial distribution functions than the Percus--Yevick (PY) and HNC closures. Furthermore, the MHNC closure is more accurate than the HNC closure even when the solute particle is nonspherical \cite{matsuo2024}. Thus, MHNC theory is best suited for the analysis in this work.

Finally, the PMFs are obtained based on the correlation functions obtained using the integral equation theories. To this purpose, the solvent--solvent correlation functions were calculated using Eq. (\ref{OZ}). Next, the Solute 1--solvent and Solute 2--solvent correlation functions are calculated using Eq. (\ref{OZ}) and Eq.  (\ref{3D-OZ}) with a common closure.
The PMF between Solutes 1 and 2 is obtained from:
\begin{equation}
    \beta W_{12}(x,y,z)=\beta u_{12}(x,y,z)-\gamma_{12}(x,y,z)-b_{12}(x,y,z),
\end{equation}
where $\gamma_{ij}(x,y,z)$ is calculated using $\hat{\gamma}_{ij}(k_x,k_y,k_z)$ given by:
\begin{equation}
\hat{\gamma}_{12}(k_x,k_y,k_z)=\sum_{i=1}^{m+2}\rho_{i} \hat{c}_{2i}(k_x,k_y,k_z)\hat{h}_{i1}(|\mathbf{k}|).
\end{equation}

\section{Results and Discussion}
\subsection{Size selectivity}
\begin{figure}[h]
 \begin{minipage}[b]{0.32\columnwidth}
\centering
  \includegraphics[width=\columnwidth]
  {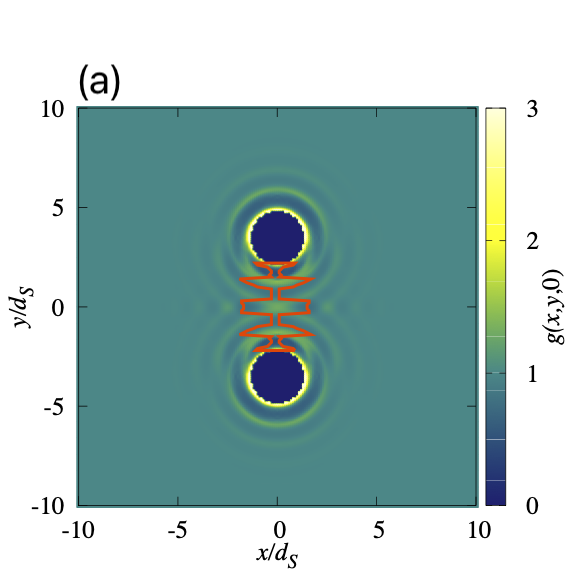}
 \end{minipage}
 \begin{minipage}[b]{0.32\columnwidth}
\centering
  \includegraphics[width=\columnwidth]
  {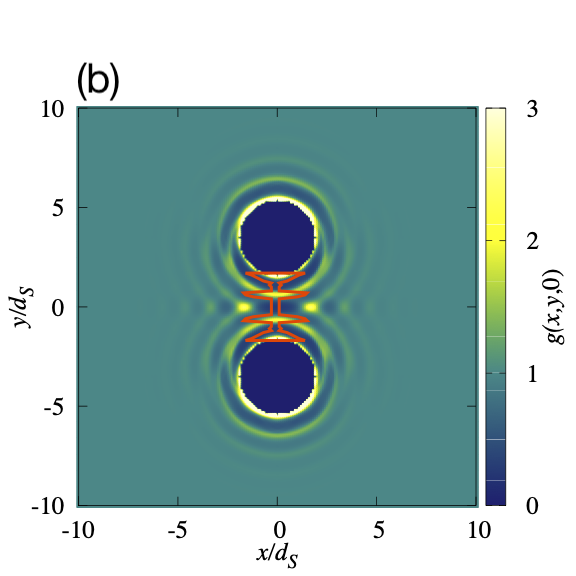}
 \end{minipage}
 \begin{minipage}[b]{0.32\columnwidth}
\centering
  \includegraphics[width=\columnwidth]
  {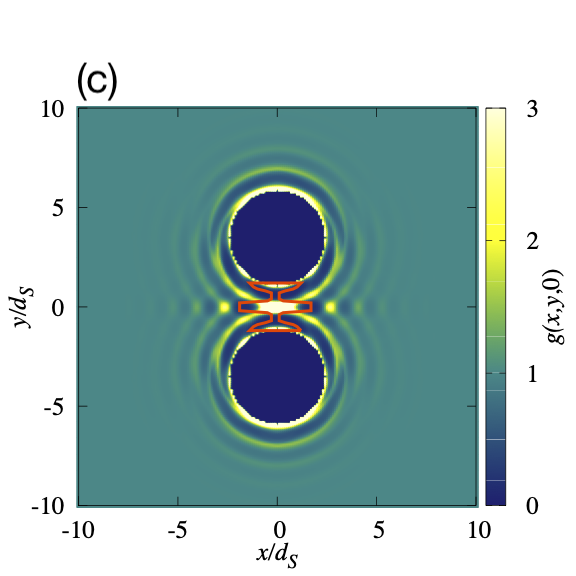}
 \end{minipage}\\
 \begin{minipage}[b]{0.32\columnwidth}
\centering
  \includegraphics[width=\columnwidth]
  {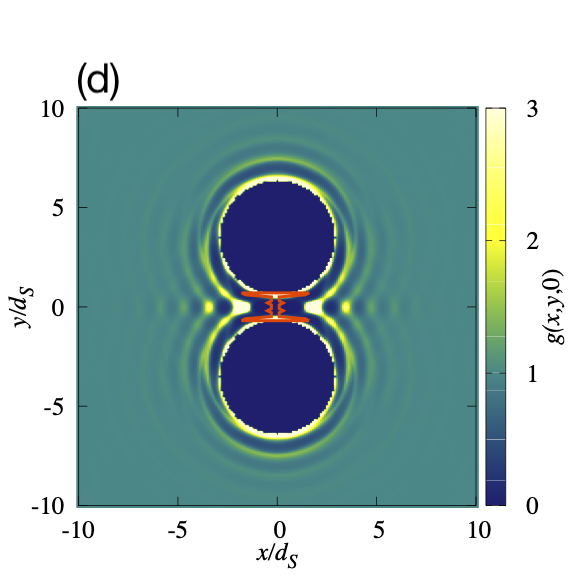}
 \end{minipage}
 \begin{minipage}[b]{0.32\columnwidth}
\centering
  \includegraphics[width=\columnwidth]
  {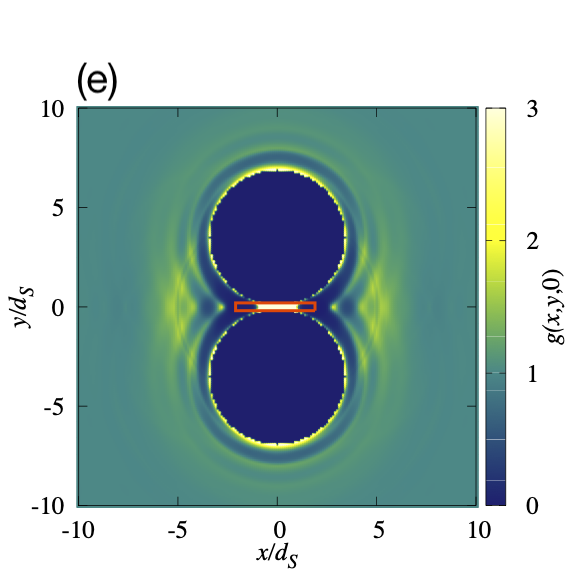}
 \end{minipage}
 \caption{The spatial distribution of guest molecules around the ring-like host in the $xy$ plane, $g_{12}(x,y,0)$, calculated using MHNC theory. The region enclosed by the red line extends from the peak position at $x=0$ to the minima along the $x$-axis. The recognition volume $v$ is defined as the volume enclosed by this region. The diameters of the guest molecules are (a) $d_S$, (b) $2d_S$, (c) $3d_S$, (d) $4d_S$, and (e) $5d_S$.}
 \label{gxz}
\end{figure}

The lock-and-key relationship is well known in host--guest affinity. The idea of entropy-driven attraction and the relationship are well-suited to each other. Kinoshita et al. adopted a simple host--guest model and showed the relationship \cite{kinoshita2002,kinoshitaoguni2002,amano2010,amano2011}. It should be confirmed that this relationship holds well in the present host--guest system because the ring-like host was not examined in those studies. We adopted the OZ–MHNC theory to confirm the validity. In the bulk calculation, Eqs. (\ref{OZ}) and (\ref{closure}) with Eq. (\ref{MHNC}) were used, and the concentration of the guest molecules ($i=1$) was finite. Then, Eqs. (\ref{3D-OZ}) and (\ref{closure}) with Eq. (\ref{MHNC}) were used to calculate the spatial distribution of the bulk around a host molecule.

The spatial distribution of guest $g_{12}(x,y,z)$ around a host molecule is presented in Fig. \ref{gxz}. We examined five sizes of candidate guest spheres with diameters $d_1=1d_S, 2d_S, 3d_S, 4d_S, 5d_S$. The host is immersed in a binary mixture, which consists of the solvent and the guest particles. The number density of the guest particles is $\rho_1=1.0\times 10^{-3}d_1^{-3}$, and the total packing fraction is kept at $0.380$. For all guest candidates, very high peaks are found around the site $(x, y, z) = (0, 0, 0)$, namely, the recognition site. The highest peak in the candidates is found for the $5d_S$ guest. These results support the validity of the lock-and-key relationship because the recognition size is almost the same as the $5d_S$ guest size.

\begin{figure}[h]
 \begin{minipage}[b]{0.43\columnwidth}
 \centering
  \includegraphics[width=\columnwidth]{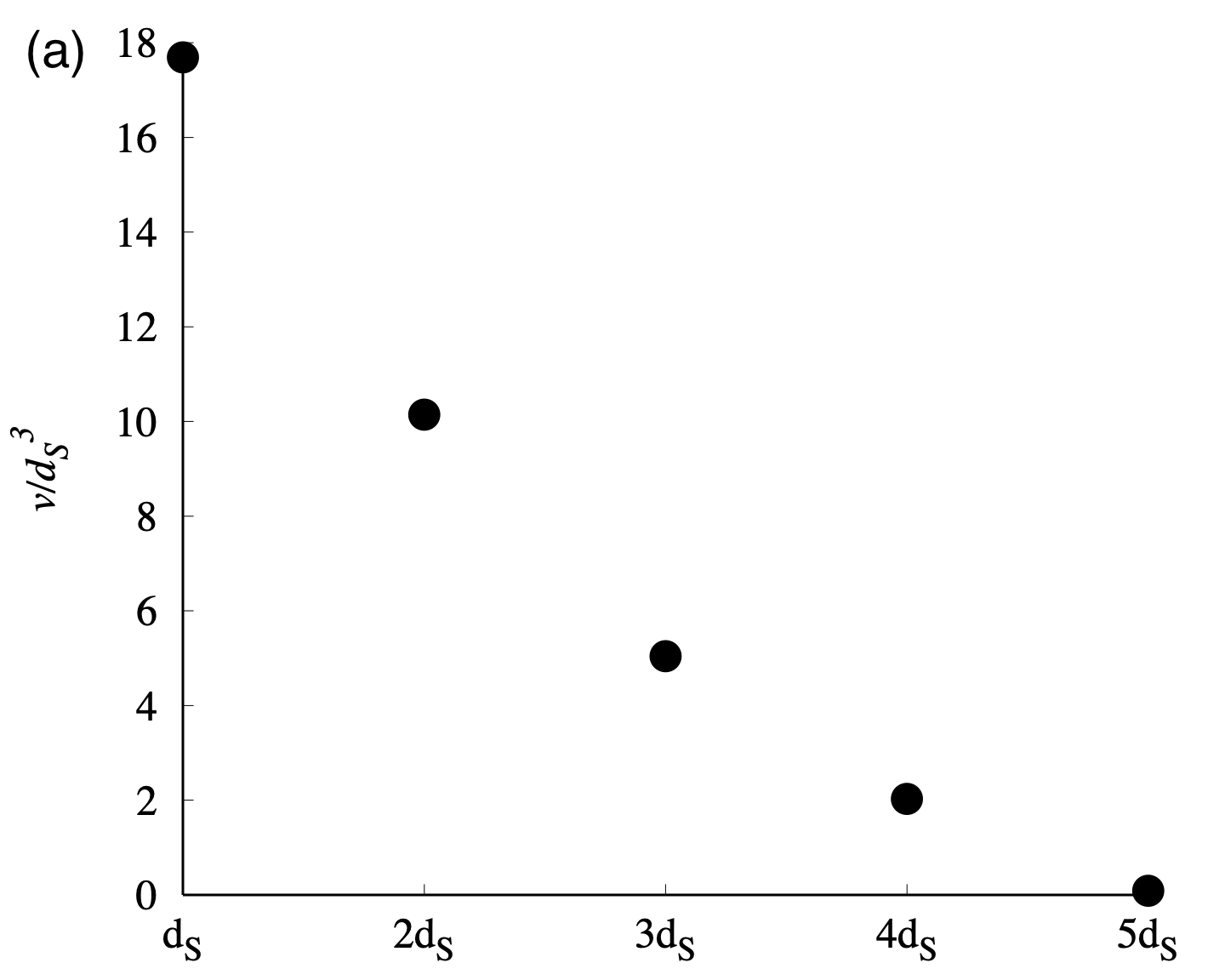}
 \end{minipage}
 \begin{minipage}[b]{0.55\columnwidth}
 \centering
  \includegraphics[width=\columnwidth]{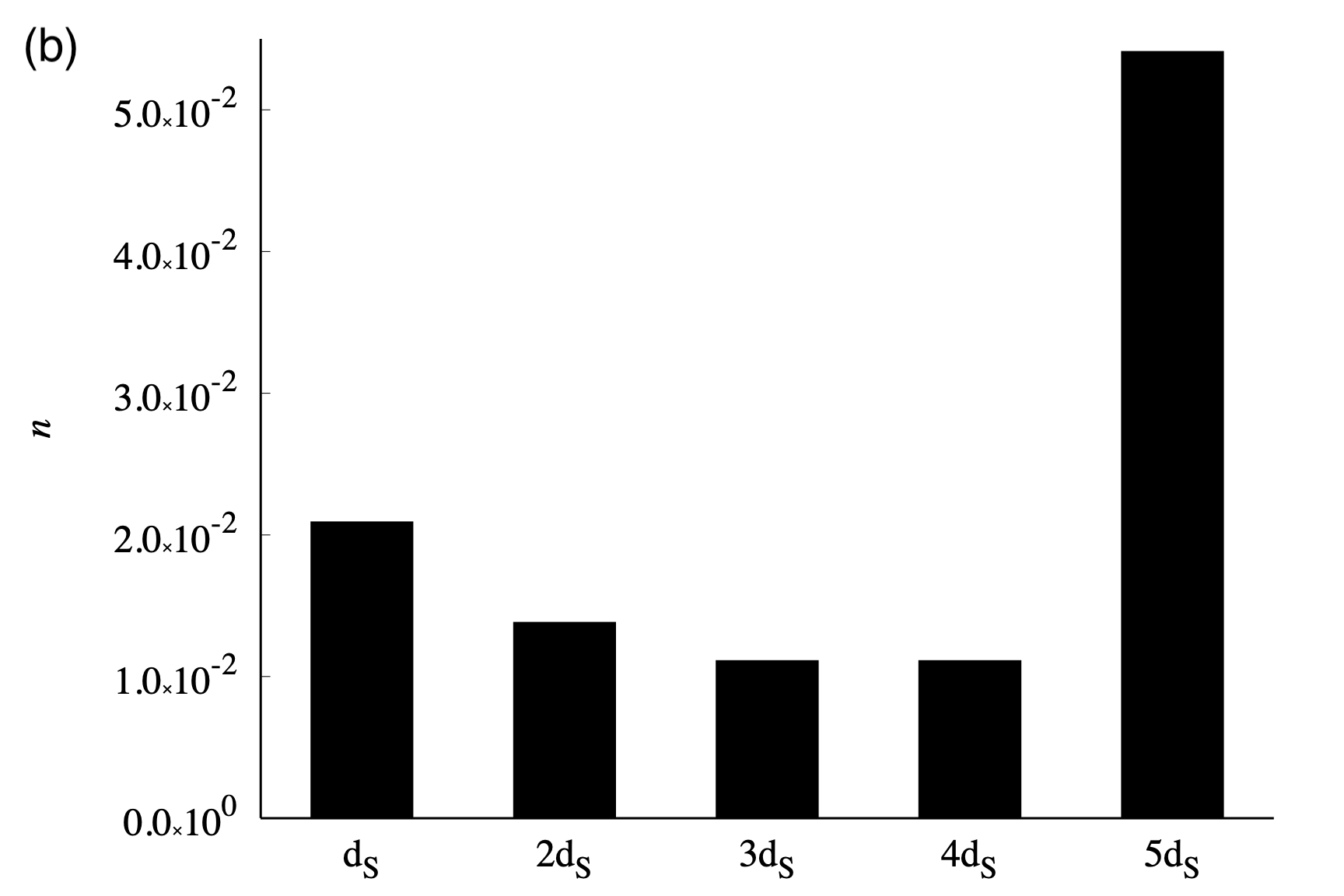}
 \end{minipage}\\
% \begin{minipage}
% \centering
%  \begin{tabular}{cccccc}
%    \hline
%    &$d_S$ & $2d_S$ & $3d_S$ & $4d_S$ & $5d_S$\\
%    \hline \hline
%    $n$ & $2.09\times10^{-2}$ & $1.38\times10^{-2}$ & $1.11\times10^{-2}$ & $1.11\times10^{-2}$ & $5.41\times10^{-2}$ \\
%    $v/d_S^3$ & 17.69 & 10.14 & 5.04 & 2.03 & 0.09\\
%    \hline
%  \end{tabular}
% \end{minipage}
\vspace*{-1cm}
\caption{(a) Recognition volume for the guest (the diameters are $d_S, 2d_S, 3d_S, 4d_S$, and $5d_S$). (b) The number of guests that the host recognizes is calculated using Eq. (\ref{num}).}
\label{selective}
\end{figure}

Next, the lock-and-key relationship is examined in terms of the number of guests the host recognizes. The necessity of this evaluation stems from the extremely small volume of the recognition site for the $5d_S$ (See Fig. \ref{selective}(a)). The number of guest particles around the recognition site was estimated by integration of the spatial distribution in the surrounding volume $v$ (red curves in Fig. \ref{gxz}). Here, $v$ is the area from the peak position at $x=0$ to the minima along the $x$-axis. We can calculate the number of recognized guest molecules as:
\begin{equation}
    n=\rho_2\iiint\limits_{v}g_{12}(x,y,z)dxdydz.
    \label{num}
\end{equation}
The value for the $5d_S$ guest is the largest among the five candidates (Fig. \ref{selective}(b)), although the recognition site volume is the smallest. This result strongly supports the lock-and-key relationship for the ring-like host.

\subsection{The PMF between host and guest molecules immersed in a one-component system}
\begin{figure}[h]
 \begin{minipage}[b]{0.49\columnwidth}
 \centering
  \includegraphics[width=\columnwidth]
  {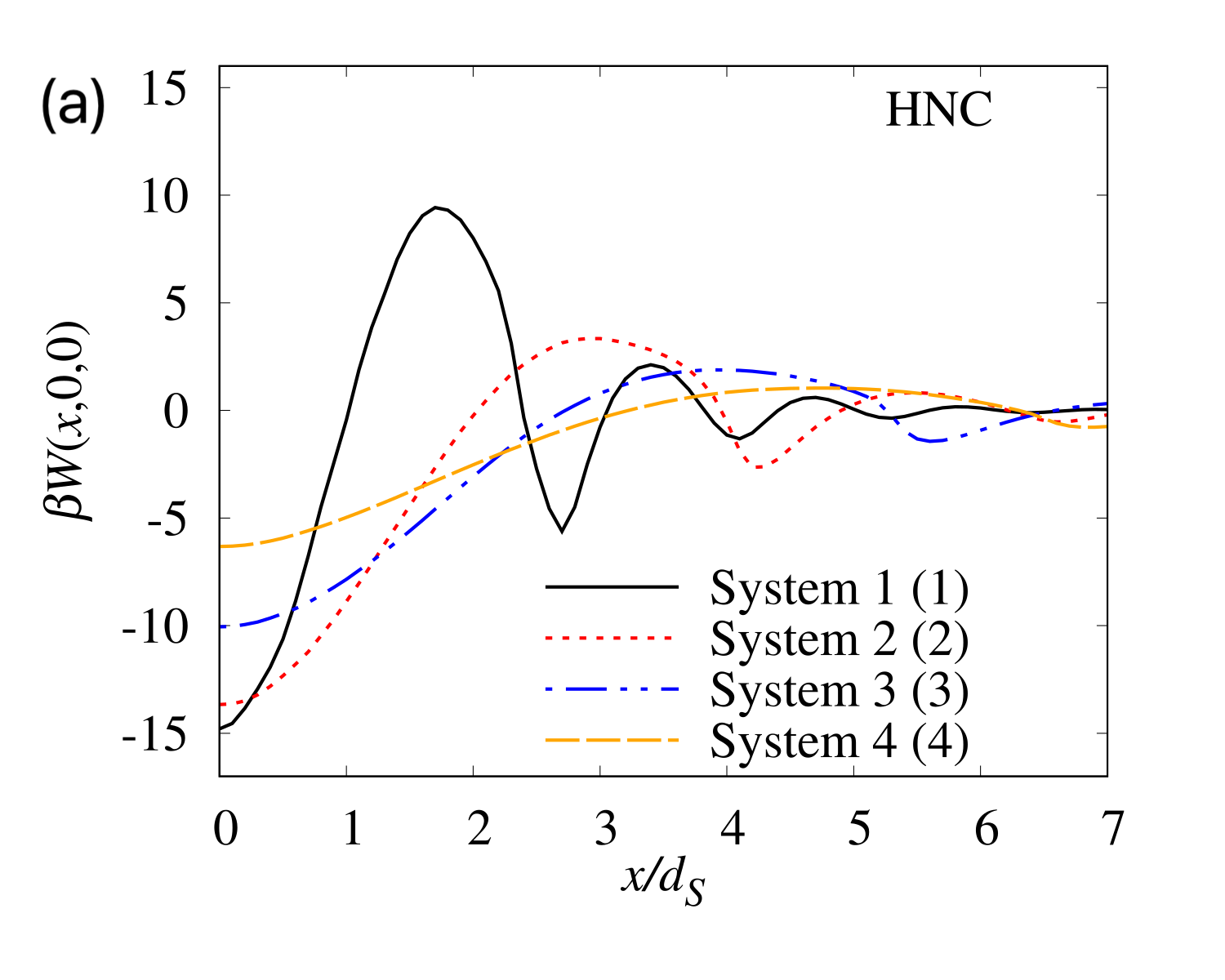}
 \end{minipage}
% \hspace{-1.5\columnwidth}
 \begin{minipage}[b]{0.49\columnwidth}
 \centering
  \includegraphics[width=\columnwidth]
  {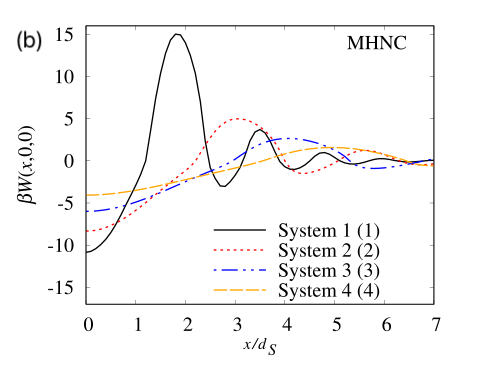}
 \end{minipage}\\
 \hspace{0.04\columnwidth}
 \begin{minipage}[b]{0.49\columnwidth}
 \centering
  \includegraphics[width=\columnwidth]
  {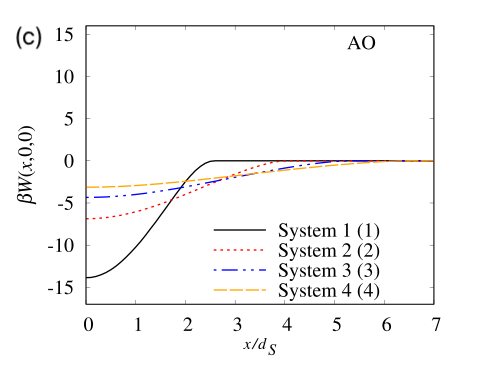}
 \end{minipage}
\vspace*{-1cm}
 \caption{PMF between the host and guest in a one-component hard-sphere fluid.  Four kinds of fluid (Systems 1--4) are examined. PMFs on the $x$-axis are shown ($x > 0$, $y = 0$, and $z = 0$). (a) HNC--OZ, (b) MHNC--OZ, and (c) AO theories.}
 \label{pz1}
\end{figure}
Two points were noted in the previous subsection. First, strong molecular recognition appears even without a direct attractive interaction between the host and the guest. The driving force is the entropy gain arising from the increase in the configurational space for the solvent particles. The increase in configurational space is equal to the reduction in excluded volume due to host--guest association. Second, the stability of the recognition for a particular host strongly depends on the guest shape. As shape matching becomes better, the effective attraction becomes stronger. And it is sensitive. When the matching becomes perfect, the stability suddenly increases. In summary, the entropic molecular recognition driven by the translational motion of solvent particles explains the so-called lock-and-key relationship well. This effective interaction must depend on the solvent size and so on. Therefore, we will discuss the effect of changes in the size of solvent molecules on the association stability of molecular recognition in this subsection.

Fig. \ref{pz1} shows the PMFs between the host and guest molecules immersed in four one-component hard-sphere fluids (Systems 1--4). Three different theories, HNC--OZ (a), MHNC--OZ (b), and AO (c), were adopted in the calculation. The PMF provides insight into the stability arising from recognition and the free-energy barrier in the recognition process. All three theories give the most stable state at the recognition site $(x, y, z) = (0, 0, 0)$ because the PMFs have the smallest values at the recognition site.

The PMFs around the recognition site, calculated using the HNC--OZ theory (Fig. \ref{pz1}(a)), show the deepest well among the three theories. When the solute particles are two hard spheres, it is known that the HNC approximation overestimates the peak value of the spatial distribution functions \cite{nakamura2019, hansen1986}. The previous study also shows that the HNC approximation overestimated peak values near a concave surface \cite{matsuo2024}. Here, the peak in the distribution function corresponded to the minimum in the PMF because of the relation $W_{ij}(x,y,z)=-k_B T \ln g_{ij}(x,y,z)$. Thus, the host--guest affinity can be estimated to be somewhat lower than the HNC results.

Fig. \ref{pz1}(b) shows the PMFs around the recognition site calculated using the MHNC--OZ theory. These results are qualitatively similar to those obtained using the HNC--OZ theory (Fig. \ref{pz1}(a)), but the MHNC results show generally weaker oscillations. Thus, the stability at the recognition site ($(x, y, z) = (0, 0, 0)$) is also lower than that obtained using the HNC--OZ theory. In our previous paper, we compared the results of Grand Canonical Monte Carlo (GCMC) simulations of the spatial distribution function near a concave surface with those of integral equation theory. The MHNC approximation slightly underestimates the peak value of the spatial distribution functions \cite{matsuo2024}. However, the MHNC results were generally closer to the GCMC results than the HNC results. We can expect the GCMC results to be closest to the exact spatial distribution function. Therefore, the MHNC results are expected to be the closest to the exact results among those obtained in the present calculation. In this paper, we primarily discuss results calculated using the MHNC--OZ theory and also refer to results from two other related theories to interpret the MHNC--OZ results.

The MHNC--OZ results (Fig. \ref{pz1}(b)) show that the free energy has a minimum when the guest particle is located at the recognition site $(x,y,z)=(0,0,0)$ for all solvent particle sizes (Systems 1--4). Moreover, the association becomes more stable as the solvent particle size decreases. These points also hold for AO theory (Fig. \ref{pz1}(c)), which ignores correlations among solvent particles. Therefore, the stability at the recognition site $(x,y,z)=(0,0,0)$ and the increased recognition stability with smaller solvent molecules can be explained in terms of excluded volume and solvent number density.

According to AO theory, the PMF between the host and guest is written as
\begin{equation}
    W_{12}(x,y,z) =-\rho k_BT \Delta V(x,y,z).
\end{equation}
Here, $\rho$ is the number density of the solvent, and $\Delta V$ represents the overlap of the excluded volume for a solvent composed of depletants. This excluded volume is the volume that the solvent particles are excluded from by the host and guest molecules. As the size of the solvent particles increases, $\Delta V$ increases. However, if the volume packing fraction is kept constant, $\rho$ decreases. Because the latter effect is dominant, the stability arising from association decreases with increasing solvent particle size. These phenomena also appear in the association process of two large rigid spheres in a rigid spherical fluid \cite{akiyama2006,rothkinoshita2006}.

By contrast, some behaviors of the stability at the recognition site $(x, y, z) = (0, 0, 0)$ are affected by correlations among solvent particles. AO theory cannot explain these effects. Here, we compare the results for System 1 (solvent particle diameter of $1d_S$) and System 2 (solvent particle diameter of $2d_S$). The MHNC--OZ results (Fig. \ref{pz1}(b)) show that the ratio of the stability for System 1 to the stability for System 2 at the recognition site $(x, y, z)=(0, 0, 0)$ is close to 1. The HNC--OZ results (Fig. \ref{pz1}(a)) also show that the ratio is almost 1. On the other hand, the results of AO theory are different. The ratio is approximately 1/2. When the diameter of the solvent particles doubles, the recognition stability is halved in the case of AO theory. Kinoshita et al. have calculated and compared the association between two large hard spheres in a hard-sphere fluid using Systems 1 and 2. Even when using HNC--OZ theory, the same destabilization of the association caused by the increase in the solvent particle size occurs as with AO theory. The solvent-particle-size dependence of this association stability, therefore, differs significantly from that observed when the guest is adsorbed onto a ring-shaped host.

Next, we discuss the oscillation behaviors of the PMF. Two integral equation theories (Fig. \ref{pz1}(a), (b)) yield PMFs with oscillations. The period of the oscillation increases as the diameter of the solvent particle becomes larger. By contrast, the PMFs calculated using AO theory do not exhibit oscillations because there is no correlation among solvent particles in this theory. Thus, those results suggest that the free-energy barriers in the recognition process arise from correlations among solvent particles. The correlation arises from short-range repulsive interactions between solvent particles at high solvent packing fractions. The short-range repulsive interaction is important when discussing the free-energy barrier in a high-packing-fraction system, such as a liquid.

The period of the oscillation shown in Fig. \ref{pz1}(a) and (b) differs from the diameter of each solvent particle. It is well known that the oscillation has a period of the solvent particle's diameter in the case of the effective interaction between large hard spheres immersed in a hard-sphere fluid. Thus, the oscillatory structure presented here seems strange. Therefore, it is necessary to explain the oscillatory behavior shown in Fig. \ref{pz1}(a) and (b) below.

The stable configuration of the ring-like host and guest occurs when the gap width between their surfaces is a natural-number multiple of the solvent diameter. Actually, we can find two concentric oscillatory structures around each sphere constituting the ring-shaped host in Fig. \ref{gxz}. The period of the spatial distribution functions is the diameter of the solvent particle. The peaks of the spatial distribution function correspond to the stable position of the effective interaction between the host particle and the solvent particle. Work is required to displace solvent particles from these stable sites. This work fluctuates periodically because the distribution function of the solvent particles is concentric. The work increases when the guest’s displacement widens the peak of this distribution function, thereby displacing the solvent particles from the stable sites. Therefore, we can expect that the guest particle is in an unstable configuration when the gap width between the two surfaces is not a natural-number multiple of the solvent particle diameter.

\begin{figure}[h]
  \centering
  \includegraphics[width=0.5\columnwidth]{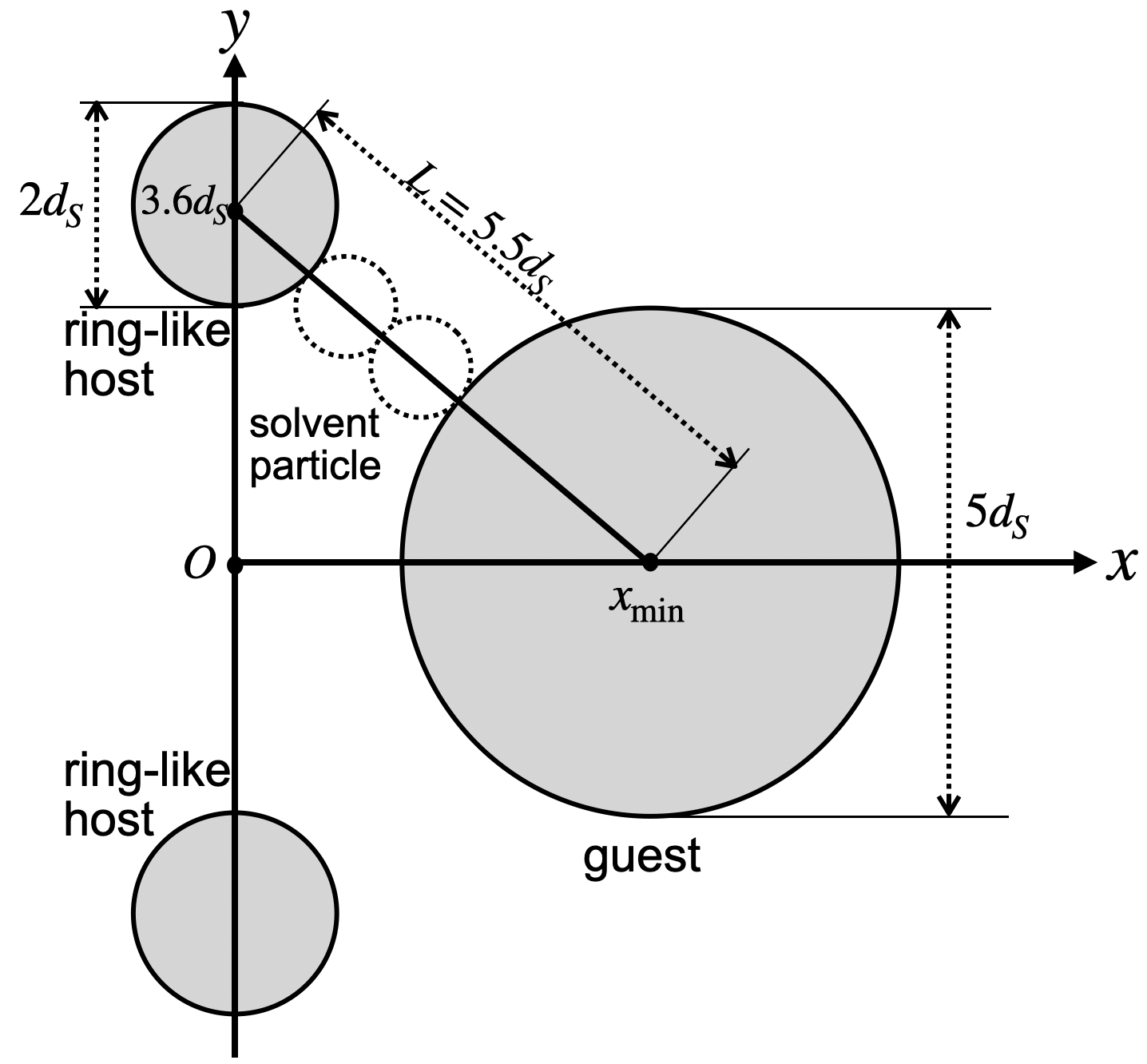}
 \caption{Schematic illustration of the host--guest arrangement used to derive Eq. (\ref{expxmin}). This configuration corresponds to the third minimum of the PMF shown in Fig. \ref{pz1}(a) and (b), where two solvent particles (diameter $d_S$) are sandwiched between the ring-like host and the guest.}
 \label{minimumposition}
\end{figure}
For example, Fig. \ref{minimumposition} shows an arrangement where two solvent particles are sandwiched between a host and a guest. In this arrangement, the distance between the center of the guest and the center of one of the spheres forming the ring-like host is $L=5.5d_S$, corresponding to the sum of the guest radius, the host-sphere radius, and the diameters of two solvent particles. Because the distance between the origin $O$ and the center of the sphere forming the ring-like host is $3.6 d_S$, and the center of the guest is located at $x_{\min}$, the Pythagorean theorem gives:
\begin{equation}
(3.6 d_S)^2 + (x_{\min})^2 =L^2= (5.5 d_S)^2.
\label{expxmin}
\end{equation}
From this relation, we obtain $x_{\min}$ in the case shown in Fig. \ref{minimumposition}.

We now generalize this idea to estimate the minimum positions of the effective interaction shown in Fig. \ref{pz1}. In this figure, the functions are drawn as the $x$-dependence of the effective interaction under the conditions $y=0, z=0$. Here, the value of $x$ at which the minimum of this dependence appears is denoted as $x_{\min}$. The value $x_{\min}$ can be inferred to appear when $n$ solvent particles (diameter $\lambda d_S$) are sandwiched between the guest (diameter $5d_S$) and one of the spheres forming the ring-like host (diameter $2d_S$), as illustrated in Fig. \ref{minimumposition}. (Here, $n$ is a natural number, and $n=2$ for the configuration shown in Fig. \ref{minimumposition}.) For $n=0$, the first minimum corresponds to $x_{\min}=0$. The relationship between $n$ and $x_{\min}$ is given by:
\begin{equation}
    (3.6d_S)^2 + (x_{\min})^2=L^2=(3.5d_S+n\lambda d_S)^2.
\end{equation}
Therefore,
\begin{equation}
    x_{\min} = [(3.5+n\lambda)^2 -3.6^2]^{1/2}d_S.
    \label{xmin}
\end{equation}
This equation rationalizes that the oscillation period shown in Fig. \ref{pz1} gradually shortens and converges to the solvent particle size as $x$ increases.

We compare the values of $x_{\min}$ with the values shown in Fig. \ref{pz1}. The above equation predicts the second minimum ($n=1$) in Systems 1--4. It also predicts the third minimum in Systems 1 and 2. Other minima for effective interactions are substantially consistent with the predictions of the above equation. However, the locations of the minima are broad, making accurate determinations difficult.
\begin{table}[]
  \caption{Values of $x_{\min}/d_S$ obtained from MHNC theory (left) and calculated using Eq. (\ref{xmin}) (right).}
  \label{tab2}
  \centering
  \begin{minipage}[t]{0.47\columnwidth}
 \centering
  \begin{tabular}{cccc}
    \hline
    & \multicolumn{3}{c}{$x_{\min}/d_S$} \\
    \cline{2-4}
    $\lambda$ & $n=1$ & $n=2$ & $n=3$ \\
    \hline
    1 & 2.8 & 4.2 & 5.4 \\
    2 & 4.4 & 6.8 &  \\
    3 & 5.8 &  &  \\
    4 & 7.1 &  &  \\
\hline
  \end{tabular}
 \end{minipage}
 \begin{minipage}[t]{0.47\columnwidth}
 \centering
  \begin{tabular}{cccc}
    \hline
    & \multicolumn{3}{c}{$x_{\min}/d_S$} \\
    \cline{2-4}
    $\lambda$ & $n=1$ & $n=2$ & $n=3$ \\
    \hline
    1 & 2.700 & 4.158 & 5.412 \\
    2 & 4.158 & 6.580 & 8.791 \\
    3 & 5.412 & 8.791 & 11.970 \\
    4 & 6.580 & 10.922 & 15.076 \\
\hline
  \end{tabular}
 \end{minipage}
\end{table}

Table \ref{tab2} summarizes the $x_{\min}$ estimated using Eq. (\ref{xmin}) and the $x_{\min}$ obtained from the PMF minimum in Fig. \ref{pz1}. When $n=1$, the agreement between them is good, but as $n$ increases, the discrepancy increases. There are two reasons for this discrepancy. First, as $n$ increases, the minimum becomes broader and shallower. This broadening makes $x_{\min}$ more sensitive to even small perturbations. Second, the stable position of solvent particles between the host and guest is also influenced by the ring-like host spheres, which are not the ring-like host sphere focused on in Fig. \ref{minimumposition}. For example, the ring-like host sphere at the bottom of the figure shifts the stable arrangement of solvent particles, which is drawn with a dotted line near the top, from the straight line connecting the center of the guest and the center of the ring-like host. Therefore, the $x_{\min}$ of PMF also deviates from the prediction by Eq. (\ref{xmin}).
\begin{figure}[h]
  \centering
  \includegraphics[width=0.8\columnwidth]{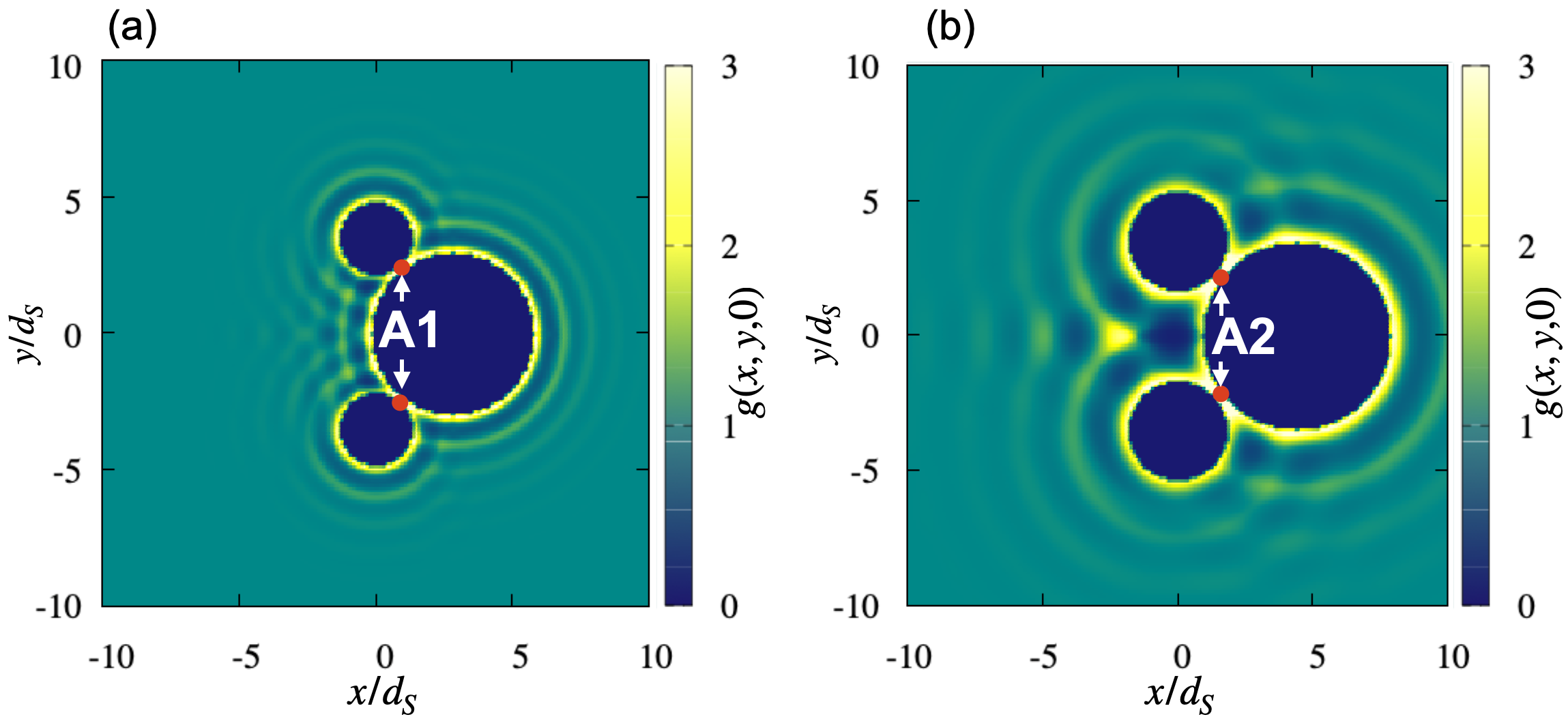}
 \caption{Spatial distributions of solvent particles around Solute 1 (guest) and Solute 2 (host) in the $xy$ plane, $g_{1-2,M}(x,y,0)$, calculated using MHNC theory. The diameters of the solvent particles are (a) $d_S$ and (b) $2d_S$. The center of the host is fixed at the origin. The center of the guest is located at (a) $x_{\min}/d_S=2.8$ and (b) $x_{\min}/d_S=4.4$. High peaks in the distribution of solvent particles appear at positions A1 and A2.}
 \label{minimummap}
\end{figure}

Fig. \ref{minimummap} shows the spatial distribution function of solvent particles when a guest molecule is placed at the second $x_{\min}$. Here, the cases for System 1 and System 2 are shown. The dark blue circular areas are the excluded volume for the solvent particles. We can find that the solvent molecules localize between the host and guest at points A1 and A2 in Fig. \ref{minimummap}. The dark blue circular areas are almost touching. The locations of the narrow gap between them are colored in a bright color. This bright color indicates the high peaks of the distribution functions, suggesting the presence of the solvent molecule at the gap. These results also support the idea discussed in the above paragraphs.

Next, we discuss how the free-energy barrier height in the molecular recognition process depends on the size of the solvent particles. Here, the nearest peak to the origin in Fig. \ref{pz1} is called the first barrier. For example, when the size of the solvent hard sphere is $1.0 d_S$, the first barrier is found near $x = 1.8 d_S$. This $x$-value lies between the first and second minima near $(0, 0, 0)$. Multiple barriers are found for any size of solvent rigid sphere, but the first barrier is the highest. Therefore, the first barrier is the most important in the recognition process.

\begin{figure}[h]
  \centering
  \includegraphics[width=0.9\columnwidth]{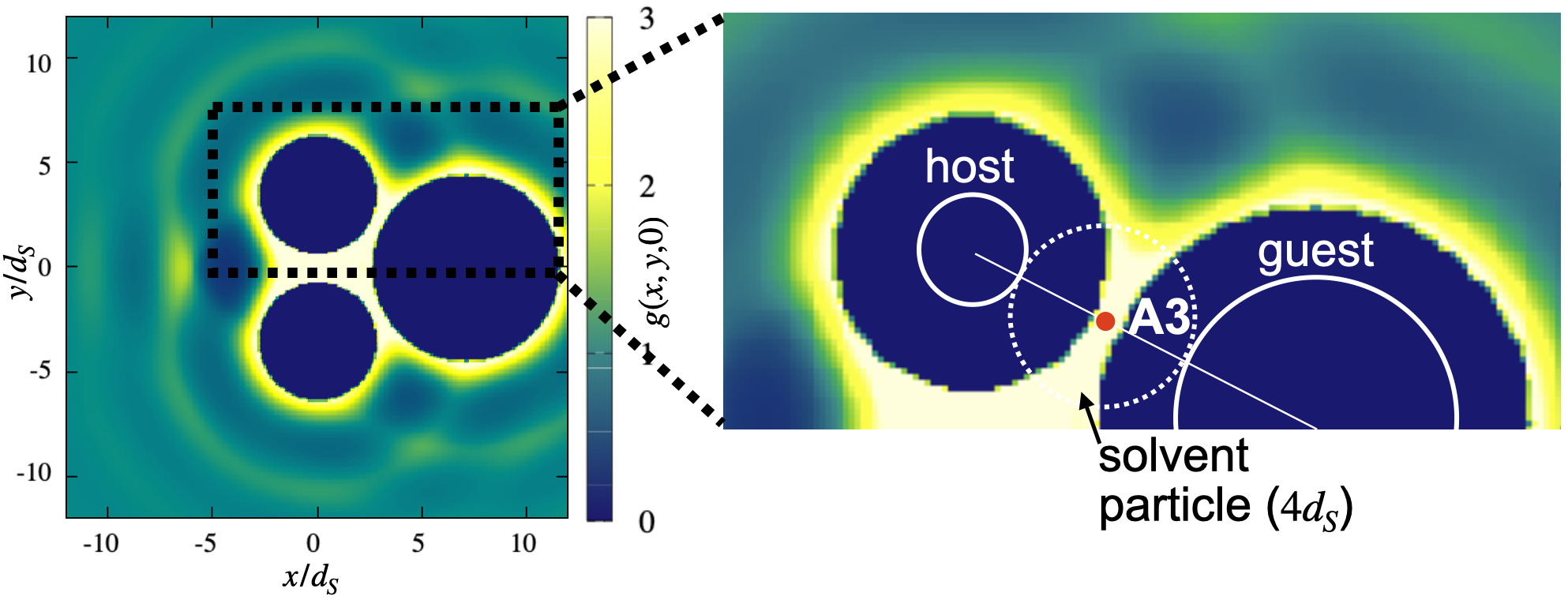}
 \caption{Spatial distribution of solvent spheres with a diameter of $4d_S$ around Solute 1 (guest) and Solute 2 (ring-like host) in the $xy$ plane, $g_{1-2,M}(x,y,0)$, calculated using MHNC theory. The center of the host is fixed at the origin. The center of the guest is located at $x_{\min}/d_S=7.1$. High peaks in the distribution of solvent particles appear at position A3.}
 \label{minimummap4}
\end{figure}

To understand this first barrier, we discuss why the effective interaction increases as $x$ decreases from the second minimum in Fig. \ref{pz1}(a) and (b). When the second minimum appears, the arrangement of the host ring and guest is as shown in Fig. \ref{minimummap4}. As shown in Fig. \ref{minimumposition}, a peak in the solvent particle distribution function appears near A in Fig. \ref{minimummap4}. That is, the gap between the host ring and the guest is the same as the diameter of the solvent particle, and a high peak of the solvent spatial distribution function appears in this gap. If there were no direct interaction between solvent particles, the solvent particles would be distributed uniformly outside the excluded volume of the host and guest. This is the particle distribution assumed in Asakura--Oosawa theory. In that case, the work was not positive when the gap narrowed (See Fig. \ref{pz1}(c)). However, in a case like Fig. \ref{minimummap4}, where there is an excess of solvent particles exerting force in the direction of widening the gap, the work done to push out solvent molecules present in the gap must be positive. Therefore, the first barrier appears between the first and the second minima in Fig. \ref{pz1}(a) and (b).

\subsection{The PMF between host and guest molecules immersed in a mixture}

\begin{figure}[h]
 \begin{minipage}[b]{0.49\columnwidth}
 \centering
  \includegraphics[width=\columnwidth]
  {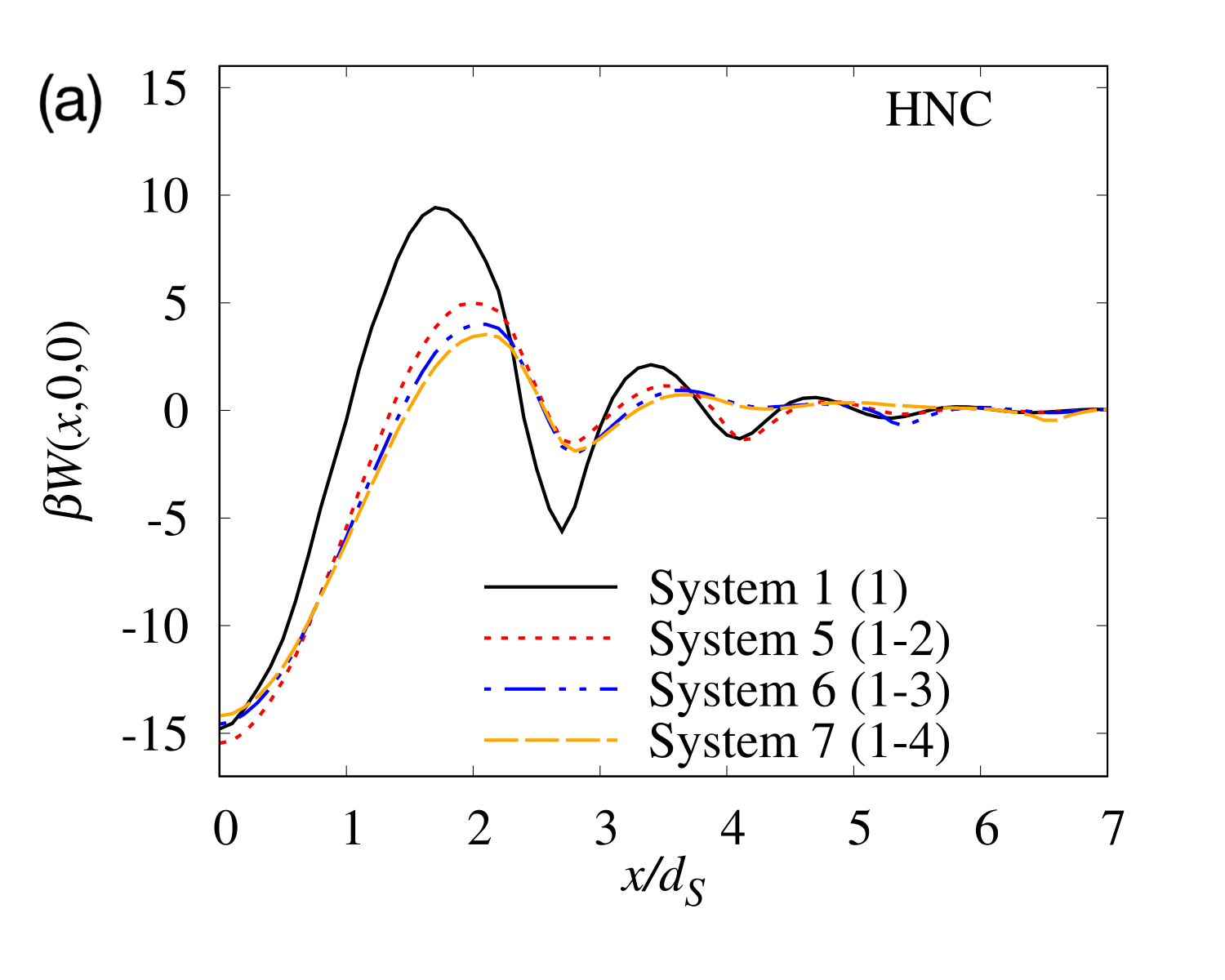}
 \end{minipage}
% \hspace{-1.5\columnwidth}
 \begin{minipage}[b]{0.49\columnwidth}
 \centering
  \includegraphics[width=\columnwidth]
  {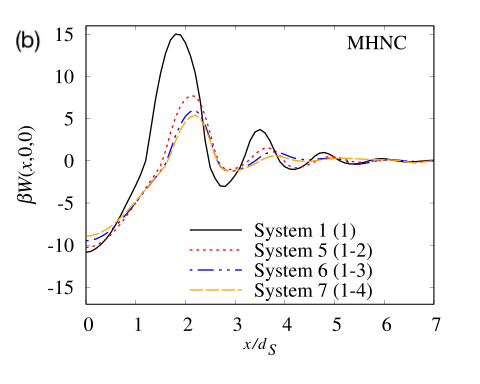}
 \end{minipage}\\
 \hspace{0.04\columnwidth}
 \begin{minipage}[b]{0.49\columnwidth}
 \centering
  \includegraphics[width=\columnwidth]
  {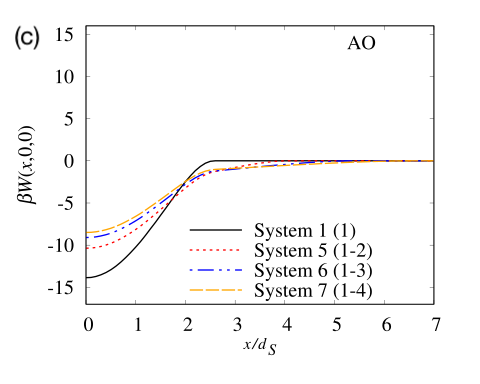}
 \end{minipage}
\vspace*{-1cm}
 \caption{The PMF between the host and guest in a one-component solvent (System 1) and binary mixtures (Systems 5--7). PMFs on the $x$-axis are shown ($x > 0$, $y = 0$, and $z = 0$). (a) HNC--OZ, (b) MHNC--OZ, and (c) AO theories.}
 \label{pz2}
\end{figure}
In the previous section, we discussed molecular recognition in a one-component hard-sphere fluid. In this section, the one-component hard-sphere fluid is replaced with a hard-sphere mixture to discuss the mixture effects. Thus, we also calculate the PMF between Solute 1 (guest) and Solute 2 (host) immersed in a hard-sphere mixture. The AO potential for the $m$-component mixture is 
\begin{equation}
    W_{12}(x,y,z)=k_B  T\sum\limits_{i=1,m+2}\rho_i\Delta V_i(x,y,z),
\end{equation}
where $\rho_i$ is the number density of the species $i$, with $\rho_1=\rho_2=0$ for the solute species, and $\Delta V_i$ is the overlap of the excluded volume for species $i$. Fig. \ref{pz2} shows the PMFs for the binary mixtures (Systems 5--7). From the results of AO theory, the stability at $x = 0$ (the first minimum) becomes lower with increasing diameter of the second component. We can again explain this dependence in terms of the solvent particle's number density. The total number density decreases when larger solvent particles are mixed under the condition of a constant total packing fraction. As discussed in the previous section, this decrease reduces stability at $x = 0$.

The decrease in stability due to the addition of large solvent particles was also observed in the effective interaction between two large hard spheres immersed in a binary hard-sphere mixture \cite{roth2006}. The destabilization was observed in the results not only by AO theory but also by HNC and MHNC theories \cite{roth2006}. However, the destabilization behavior only appears in the results using AO theory for the present host--guest effective interaction. In Fig. \ref{pz2}(a) and (b), the system dependence at the first minimum is much smaller than that in Fig. \ref{pz2}(c). The orders of the stability in Fig. \ref{pz2}(a) and (b) are different from those in Fig. \ref{pz2}(c). Based on our previous research and the discussion in the previous section, the following discussion will mainly focus on the results of MHNC theory, with the results of HNC theory presented as supplementary information.

Next, we discuss the height of the first barrier. The heights obtained using MHNC theory for Systems 5--7 are smaller than the barrier for System 1 (Fig. \ref{pz2}(b)). However, the height weakly depends on the diameter of the second component. Moreover, the effect of the second component on the period of PMF oscillation was also weak. The PMF period in Systems 5--7 was close to $d_S$ (i.e., the diameter of the first component). Because the number density of the first component was much larger than that of the second component, the effect of the first component on the PMF oscillation period was dominant. As a result of this effect, the barrier heights for Systems 5--7 were nearly identical.

\begin{figure}[h]
 \begin{minipage}[b]{0.49\columnwidth}
 \centering
  \includegraphics[width=\columnwidth]
  {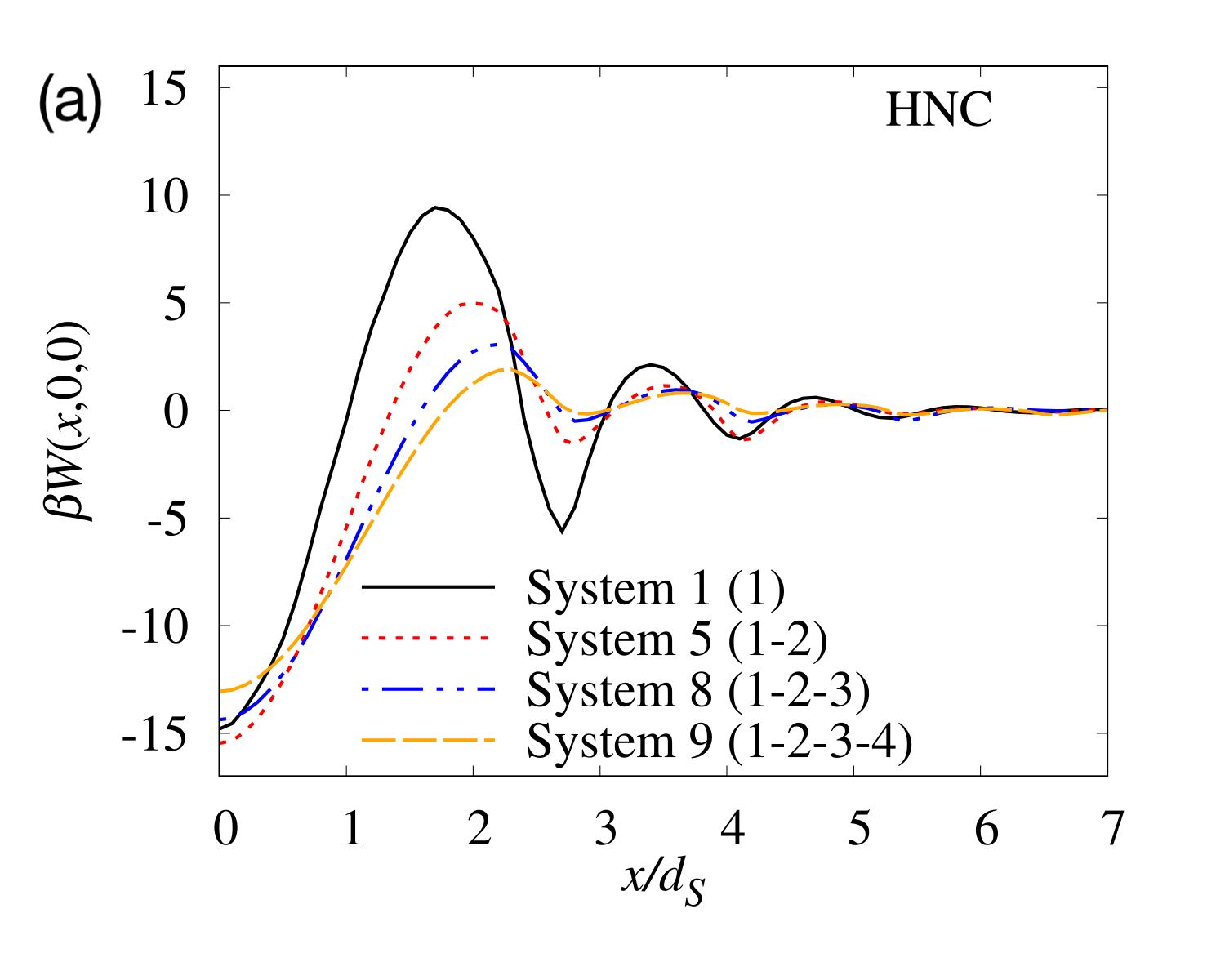}
 \end{minipage}
% \hspace{-1.5\columnwidth}
 \begin{minipage}[b]{0.49\columnwidth}
 \centering
  \includegraphics[width=\columnwidth]
  {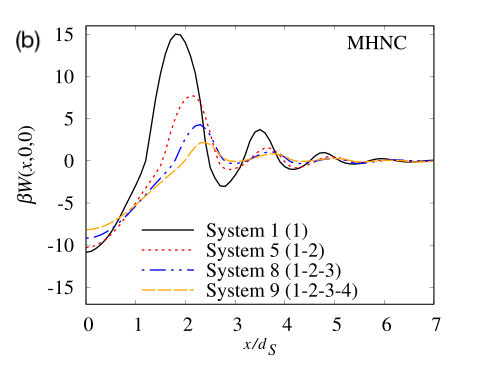}
 \end{minipage}\\
 \hspace{0.04\columnwidth}
 \begin{minipage}[b]{0.49\columnwidth}
 \centering
  \includegraphics[width=\columnwidth]
  {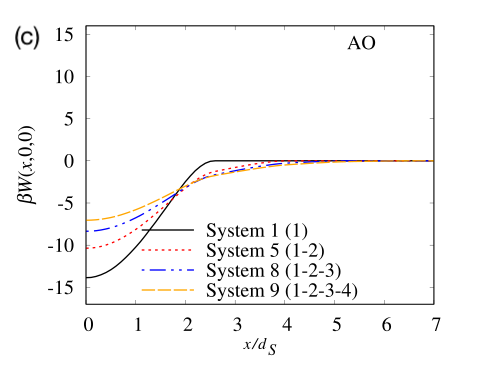}
 \end{minipage}
\vspace*{-1cm}
 \caption{PMF between the host and guest in one-, two-, three-, and four-component solvents (Systems 1, 5, 8, and 9). PMFs on the $x$-axis are shown ($x > 0$, $y = 0$, and $z = 0$). (a) HNC--OZ, (b) MHNC--OZ, and (c) AO theories.}
 \label{pz3}
\end{figure}

As shown above, the addition of the second-component solvent particles does not reduce the height of the first peak. However, the addition of some other components reduced the barrier height in the case of the PMF between two large hard spheres \cite{roth2006}. Therefore, we examined a similar strategy to reduce the first barrier in the present system: ring-like and spherical guests immersed in a hard-sphere fluid. The PMFs for Systems 8 and 9 are shown in Fig. \ref{pz3}. Systems 8 and 9 have three- and four-component solvent particles, respectively. The free-energy barriers in these multicomponent systems were lower than those in the binary mixtures and are much lower than those in System 1. As the variety of the sphere diameters (i.e., $d_S$--$3d_S$ and $d_S$--$4d_S$ in Systems 8 and 9, respectively) increases, the barrier becomes lower.

Here, we compare the depth of the first minimum in System 1 and that in System 9 numerically again. In AO theory, the reduction in total number density led to a decrease in System 9's stability. The value for System 9 was almost half ($48\%$) that of System 1. By contrast, the results for MHNC theory are different. The reduction is much smaller than that obtained by AO theory. In MHNC theory, the stability in System 9 is $-8.16k_BT$, which is smaller than that in System 1 ($-10.85k_BT$) by only $25\%$. The correlation between solvent particles causes this difference. In AO theory, the correlation is ignored, and the reduction in the total number density directly affects recognition stability. The packing around the host, guest, and surrounding solvent particles becomes important if correlations between the solvent particles exist. Multibody effects cause these nontrivial phenomena. These behaviors of the first minimum are similar to those calculated using HNC theory. In our previous study [Matsuo 2024], the peak of the spatial distribution near a concave surface, calculated using GCMC, is intermediate between the results obtained using HNC theory and MHNC theory.  Therefore, we can expect similar behaviors in the GCMC result.

Here, let us focus on the region near the center of the ring, i.e., around $(x,y,z)=(0,0,0)$. In the case of particles smaller than $4d_S$, the particles are relatively excluded from that region as the particle size increases (See Fig. \ref{selective}). Therefore, the influence of $4d_S$ and $3d_S$ particles on the stability of guest particles around $(0,0,0)$ may not be very large. The discussion based on the partial molar volume change must be important in this recognition process. The weak dependence on the first well depth should be useful in the application. These phenomena must be confirmed using a GCMC simulation as well.

First, the barrier difference is discussed based on the MHNC results. Here, we numerically compare the height of the first barrier in System 1 with that in System 9. The first free-energy barrier in System 9 was $2.16 k_BT$, about $14\%$ of that for System 1 ($15.02 k_BT$). This dramatic reduction in the highest barrier in the recognition process is useful in practice for controlling the reaction rate when solvent mixing is easy.

\begin{figure}[h]
 \centering
 \begin{minipage}[t]{0.47\columnwidth}
 \centering
  \includegraphics[width=\columnwidth]
  {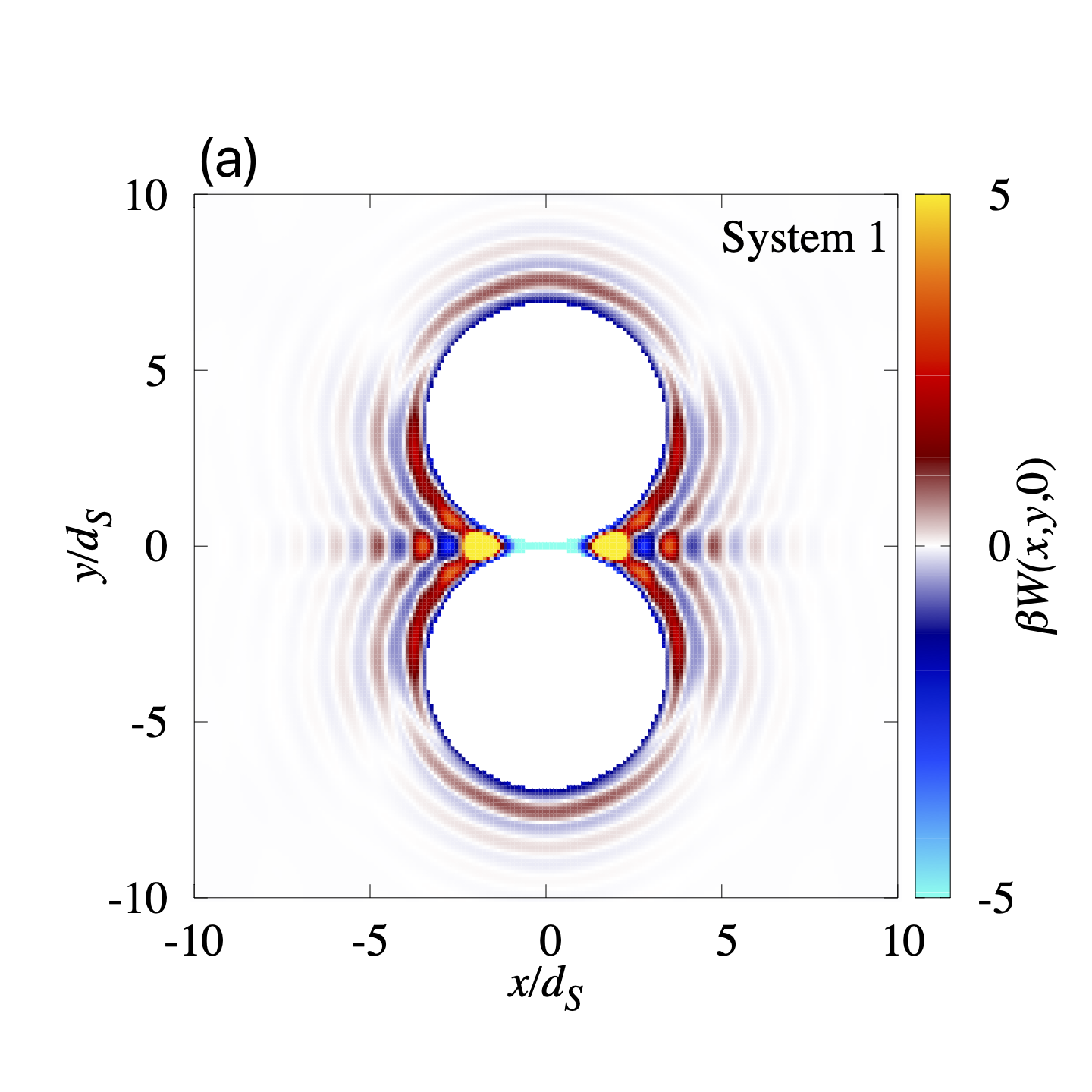}
 \end{minipage}
 \begin{minipage}[t]{0.47\columnwidth}
 \centering
  \includegraphics[width=\columnwidth]
  {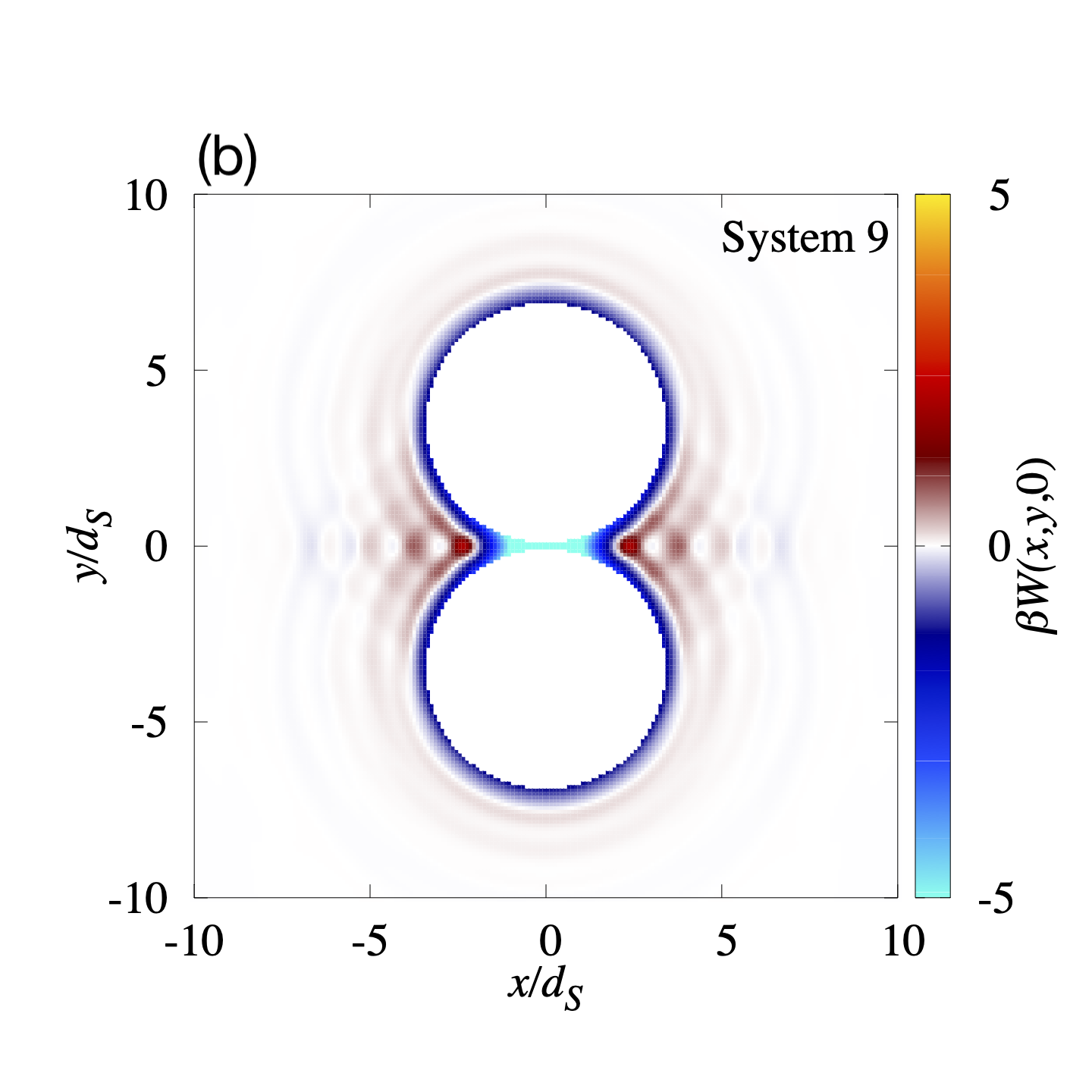}
  \label{1234pxz}
 \end{minipage}
\vspace*{-1cm}
\caption{PMF on the $xy$ plane in (a) System 1 and (b) System 9 obtained using MHNC theory.}
\label{pxz}
\end{figure}

The above results for the first barrier on the solvent mixing effect are nontrivial. Here, we rationalize this height reduction due to solvent mixing by focusing on the interference of static density waves in the spatial distribution. The spatial distributions of the PMF for Solute 1 (guest) around Solute 2 (host) on the $xy$ plane are drawn in Fig. \ref{pxz}. We set the host's ring center at the origin. The two vertically aligned white circular regions represent regions where the guest cannot place because of the spheres that constitute the host. Maps (a) and (b) are the PMFs for Systems 1 and 9, respectively. These PMFs are calculated using MHNC theory under the dilute-guest condition. Because there is a relation $g({\mathbf{r}})=\exp[-\beta W({\bf r})]$, both maps show that the fixed host is the source of the static density wave of the guest.

%\begin{figure}[b]
% \centering
%  \includegraphics[width=0.7\columnwidth]
%  {interference.png}
%  \caption{Picture illustrating density wave interference in a mixture.}
%\label{interference}
%\end{figure}
We can confirm the above picture in System 1 (Fig. \ref{pxz}(a)). The solvent in this system is a single component with a diameter of $1 d_S$. As explained previously, the $x$-axis oscillation period of the effective host--guest interaction in System 1 is not $1d_S$. (See Figs. \ref{pz2} and \ref{pz3}.) However, in the normal direction of the wavefront, the period is $1 d_S$. (See Fig. \ref{pxz}(a).) Moreover, interference patterns appear between the two fixed spheres. If the superposition principle holds, the combined effective interaction can be described as the sum of the effective interactions. In the present case, if the superposition principle held, the effective host--guest interaction could be obtained by simply summing up the effective interactions between the guest and each of the 12 spheres constituting the ring-host. As shown in previous papers \cite{matsuo2024,kubota2012}, the superposition principle on the static density wave in the spatial distribution functions is not rigorous, but it is qualitatively correct. Fig. \ref{pxz}(a) also suggests that the superposition principle qualitatively holds.

We can explain the disappearance of the highest barrier in the host--guest effective interaction in System 9 by considering one more point. This point concerns interference between waves with different periods. This destructive interference is shown in the case of an effective interaction between two large hard spheres in a hard-sphere mixture \cite{rothkinoshita2006}. We can estimate that the effective host--guest interaction is approximately equal to the sum of the effective interactions between the guest and each of the 12 spheres that constitute the ring-host. Thus, the first barrier disappears in System 9. The superposition principle is qualitatively useful in considering strategies to smooth the host--guest association processes.

\section{Conclusion}
The PMFs between a ring-like model molecule and a large sphere in a multicomponent mixture of smaller spheres were investigated using AO, OZ--HNC, and OZ--MHNC theories. AO theory is the simplest theory to describe the depletion interaction. The interaction between small spheres is ignored in AO theory; therefore, the predicted depletion interactions do not oscillate. By contrast, the real entropic force is oscillatory and possesses a free-energy barrier, which plays an important role. Nine kinds of hard-sphere fluids were prepared, and the stabilities of the recognition and the free-energy barrier in the recognition process were obtained.

The PMFs in the mixtures differed significantly from those in a one-component system. The oscillatory structure of the depletion interaction in a multicomponent system was lower than in a one-component system, and the free-energy barrier for the association process in a multicomponent system was reduced. The strongest reduction of the free-energy barrier was observed in a multicomponent mixture of smaller spheres with varying diameters, with the same packing-fraction value of each component. This feature was similar to the previous study of spherical solutes \cite{rothkinoshita2006}. When the viscosities of the solvents are similar to each other, this could mean a reduction in the time scale in which the guest molecules reach a recognition site of host molecules. Furthermore, the PMF in the multicomponent mixture has a short-range attraction in the cavity, which induces sufficient stability for molecular recognition.

In the case of spherical solute particles, the stabilities were reduced when the barrier heights were reduced in the previous study by Kinoshita and Roth \cite{rothkinoshita2006}. By contrast, in the present study, the stabilities were maintained when the barrier heights were reduced. These results showed that the free-energy barrier could be controlled while the recognition maintained sufficient stability by choosing an appropriate multicomponent mixture. However, the approximation of the integral equation is not perfect. In particular, previous studies have shown that the approximation near concave surfaces was worse than that near convex surfaces \cite{matsuo2024}. We should verify it with a GCMC simulation using the present model.

\section{Acknowledgments}
This work was supported by the Japan Society for the Promotion of Science (JSPS) KAKENHI (Grant Nos. JP23K04523, JP22K03558, JP21K18604, JP19H01863, and JP19K03772) and JST, the establishment of university fellowships toward the creation of science technology innovation, Grant Number JPMJFS2132. The computations were performed using the Research Center for Computational Science, Okazaki, Japan (Project: 23-IMS-C088, 24-IMS-C083, 25-IMS-C222) and the Research Institute for Information Technology, Kyushu University.
\bibliography{ref}
\bibliographystyle{elsarticle-num}
\end{document}